\documentclass[journal]{IEEEtran}
\usepackage{amsmath,amsfonts,amssymb,amsthm,bbm}
\usepackage{soul}
\usepackage{algorithm}
\usepackage{array}
\usepackage[caption=false,font=normalsize,labelfont=sf,textfont=sf]{subfig}
\usepackage{stfloats}
\usepackage{graphicx}
\usepackage{cite}
\usepackage{verbatim}
\usepackage{tikz}
\usepackage{xcolor}
\usepackage{algpseudocode}

\renewcommand\vec{\mathbf}
\newcommand{\mat}[1]{\mathbf{#1}}
\usepackage{soul}
\usepackage[normalem]{ulem}

\usepackage{float}
\newfloat{optim_prob}{t}{prob}
\floatname{optim_prob}{Problem}
\usepackage{caption}
\allowdisplaybreaks[4]

\begin{document}

\title{Spectrum and RAN Sharing: How to Avoid Cross-Subsidization While Taking Full Advantage of Massive MU-MIMO?}

\author{\IEEEauthorblockN{Abdalla Hussein, Patrick Mitran and Catherine Rosenberg}
\thanks{The authors are with the Department of Electrical and Computer Engineering, University of Waterloo, Waterloo, ON N2L 3G1, Canada (e-mails: \{am3abdelaziz,pmitran,cath\}@uwaterloo.ca). }
}

\maketitle

\begin{abstract}
Motivated by the need to use spectrum more efficiently, this paper investigates fine grained spectrum sharing (FGSS)
in Multi-User massive MIMO (MU-mMIMO) systems 
where a neutral host enables users from different operators to share the same resource blocks. To be accepted by operators, FGSS must i) guarantee isolation so that the load of one operator does not impact the performance of another, and ii) avoid cross-subsidization whereby one operator gains more from sharing than another.

We first formulate and solve an offline problem to assess the potential performance gains of FGSS with respect to the static spectrum sharing case, where  operators have fixed separate sub-bands, and find that the gains can be significant, motivating the development for online solutions for FGSS.
Transitioning from an offline to an online study presents unique challenges, including the lack of apriori knowledge regarding the performance of the fixed sharing case that is required to ensure isolation and cross-subsidization avoidance. We overcome these challenges and propose an online algorithm that is fast and significantly outperforms the static case.

The main finding is that  FGSS for a 
MU-mMIMO downlink system is doable in a way that is ``safe" to operators and brings large gains  in spectrum efficiency (e.g., for 4 operators, a gain above  60\% is seen in many cases).
\end{abstract}


\section{Introduction}
Spectrum with good propagation characteristics is scarce and in high demand.
Many approaches can be used to improve spectral efficiency. 
One such technique is spatial reuse through the densification of a cellular network. Nevertheless, densification provides diminishing returns and can be costly. Another approach is to use transmission techniques such as multi-user massive MIMO (MU-mMIMO) whose benefits have been studied extensively. Yet another approach is for cellular operators to  share their spectrum in a fine grained manner that is more flexible than 
current practice.
%currently done. 
This paper investigates the latter approach. Specifically, this paper develops an online method for inter-operator flexible fine grained spectrum  sharing  (FGSS). 

FGSS is made possible by two recent trends, namely Radio Access Network (RAN) sharing  and MU-mMIMO. 
While traditionally the only physical infrastructure that was shared between operators were towers and  utilities, the advent of RAN sharing and slicing has enabled spectrum sharing at the resource block level.
RAN sharing is typically performed by having an entity, called Neutral Host (NH), manage the infrastructure on behalf of a set of competing operators. The spectrum then either belongs to the NH who apportions it to the operators based on individual service level agreements (SLAs) or each operator brings its own spectrum. In both cases, the total spectrum can be shared in different ways. 

In the traditional RAN sharing approach,  a static spectrum allocation paradigm is used. Each operator is allocated a slice with a fixed portion of the total spectrum, corresponding to the amount bought or brought by the operator. 
This method guarantees slice isolation, i.e., the performance of each slice depends solely on its own load and its spectrum share.
This will be the \emph{benchmark} against which FGSS will be compared.
 
With the emergence of MU-mMIMO, it now becomes possible to multiplex several users on a single Physical Resource Block (PRB), the smallest unit of resource to allocate (see Section~\ref{sec:System_model} for more details). 
With static spectrum sharing (i.e., the benchmark) the pool of users that can be multiplexed on a PRB is limited to the users of the operator that owns the PRB. In contrast, with FGSS, the pool of users is increased to all users of all operators, significantly increasing multiplexing opportunities.
Indeed, we will show that the benchmark can be very inefficient compared to FGSS. However, for operators to accept FGSS, FGSS methods must be developed that i) provide isolation to each operator (which we define precisely in this case as the performance seen by an operator is never worse than the one it would receive in the benchmark) 
and ii) avoid cross-subsidization whereby one operator gains more than another from sharing.

This paper develops such an FGSS method for  a single cell downlink MU-mMIMO Orthogonal Frequency Division Multiple Access (OFDMA) system when Zero-Forcing Transmission (ZFT) is used. 

The main contributions of this work are:
\begin{enumerate}
    \item We formulate a mathematical optimization problem for FGSS over an OFDMA frame that aims to maximize the gain (with respect to the static sharing benchmark) achieved per slice while ensuring isolation and cross-subsidization avoidance, i.e., equal gain per slice. This problem applies to any number of non-identical slices, i.e., to operators with different shares of the common band. Note that  i) this problem requires as an input the performance each slice would receive in the frame if the benchmark, i.e., static spectrum sharing, was used; ii) if cross-subsidization avoidance is ensured and the gain is greater than one, then isolation is achieved. 
    \item 
    The above FGSS problem is a large mixed-integer non-linear problem, and cannot be solved directly. 
    Instead, we propose a dynamic sequential algorithm for solving the formulated FGSS problem. Specifically, we propose to solve a sequence of per PRB weighted sum-rate maximization problems, where the weights are changed every PRB to reflect the priority of each user and each slice in receiving resources to maximize performance and minimize cross-subsidization, respectively.
   \item To evaluate the potential gains of using the sequential algorithm for FGSS, we first consider an offline setting for solving the resulting weighted sum-rate problems, and take the static benchmark performance as known. We find that for this offline setting, all slices perform significantly better than in the benchmark, isolation is achieved and the gains are quasi-constant across slices. The gains are significant and  can be as high as 71\% (resp. 193\%) when the number of slices is 2 (resp. 4).   
    \item To  make the proposed method practical and online, we first need to address the issue of requiring the static benchmark performance at the start of each frame. For this, we propose using a Graph Neural Network (GNN) to predict the per slice benchmark performance on a per frame basis using  the large scale fading of each user.
    We show that the proposed GNN is capable of accurately predicting the performance with a mean absolute percentage error of less than 5\% for a wide range of configurations, i.e., number of users per slice and resource share per slice.   
    \item Finally, we propose a simplified method that reduces the computational complexity required for solving the weighted sum-rate problem in each PRB. The innovation is  an efficient and quick heuristic for user-selection. Combining it
    with the GNN that estimates the static benchmark performance, we obtain an online FGSS scheme that achieves nearly the same performance as the offline scheme across a wide-range of numbers of users, slices, and deployment scenarios.     
\end{enumerate}

\noindent {\bf Main Finding}: Fine grained spectrum sharing for the downlink of a MU-mMIMO system not only brings significant gains in spectrum efficiency but can be performed in a way that is ``safe" to operators by avoiding cross-subsidization.

\vspace{6pt}

The remainder of the paper is organized as follows: Section~\ref{sec:Review} reviews related works. Section~\ref{sec:System_model} provides a detailed overview of the system model. 
The formulation of FGSS and the proposed sequential  approach are presented in Section~\ref{sec:DRS}.
In Section~\ref{sec:online_solution}, we present a complete online scheme that combines two innovative components: a GNN to predict the benchmark performance at the start of each frame, and an efficient heuristic for user-selection suitable for quick online computations. Section~\ref{Sec:results} presents the simulation results and performance analysis of the proposed offline and online solutions. Finally, Section~\ref{sec:conclusion} concludes the paper.

\emph{Notation:} matrices and vectors are set in upper and lower boldface, respectively. The operators $(.)^T$, $(.)^H$, $(.)^\dagger$ and $||.||$ denote, the transpose, the Hermitian transpose, the pseudo inverse and the Euclidean norm. Calligraphic letters, such as $\mathcal{A}$, denote sets and $|\mathcal{A}|$ denotes the cardinality of set $\mathcal{A}$. 
Finally, $\mathbb{C}$ denotes the complex numbers.

%===================================================================================
    \section{Related Work } \label{sec:Review}
%===================================================================================

Network slicing has been introduced in  5G as a means to promote sharing of the underlying physical network infrastructure among entities such as Virtual Network Operators (VNOs). Radio Access Network (RAN) slicing is the process of dividing the radio resource, such as power and sub-channels, between slices such that the Service Level Agreement (SLA) between the network operator and each slice owner is enforced. Although the defacto approach of allocating fixed radio resources to each slice provides high performance isolation, it leads to over-provisioning of radio resources, which is costly, especially in a multi-user transmission setting. 
While fine grained radio resource sharing is necessary for the efficient use of radio resources, it needs to be done in a way that provides  isolation and cross-subsidization avoidance.

The definition of an SLA is an open question that greatly depends on the slicing use-case which dictates the requirements to be enforced. For example, if each slice provides a specific service to its users, referred to as slicing-as-a-service, the SLA will only care about providing an adequate level of Quality of Service (QoS) while providing an efficient usage of the available resources. On the other hand, when slicing is used in a Neutral Host (NH) use-case, then as the slices are competing VNOs, the SLA should ensure equal gain to all slices while sharing spectrum efficiently.

Several slicing solutions have been developed for the single-user transmission setting, i.e., when the base-station can serve at most one user per PRB. To improve the overall spectrum utilization, a two-level scheduler that first gives PRBs to slices, then distributes these resources among users has been proposed in \cite{ferrus20185g,da2016impact,foukas2017network,costa2013radio}. Several studies proposed solutions for the top level scheduler \cite{sciancalepore2017mobile,mandelli2019satisfying}.  A joint-scheduler based on deep reinforcement learning was proposed in \cite{khodapanah2020slice}. 

Ensuring isolation, or providing so-called performance-based SLAs, has also been investigated for single-user transmission systems. The main methodologies rely on control methods for correcting resource allocation based on the deviations from the target performance measures, as in \cite{khodapanah2018fulfillment}. Alternatively, in \cite{caballero2018network} and \cite{caballero2019network}, the problem of maintaining isolation was formulated as a game where slices trade resources in an attempt to do better, and thus each slice is responsible for choosing the right actions to maximize its own performance. 

In \cite{joshi2017dynamic}, dynamic sharing for MU-MIMO was proposed for the case of two equally sized VNOs in a single cell neutral host system. The sharing is done under a pricing scheme and each VNO is assumed to have a constraint on how much they are willing to pay at any time instant. The objective of the system is to find an optimum bargain between the operator and the slice owners without violating the packet delay constraints. However, the authors neither consider the general case of arbitrary sized slices, cross-subsidization avoidance, nor have they considered the full complexity of massive MIMO transmission, including power-allocation across multiple channels, and modulation and coding scheme (MCS) selection.

Despite the existence of several works on RAN slicing in the literature, almost none have focused on multi-user transmission and none have considered cross-subsidization avoidance, which is required for RAN sharing to be accepted by operators. 

%===================================================================================
\section{System Description and Assumptions} \label{sec:System_model}
%===================================================================================
In this section, we present the system model for the downlink of a single-cell Neutral Host (NH) system hosting $Q$ operators, also referred to as tenants. In other words, $Q$ operators are sharing a common RAN. The base station is equipped with a MU-mMIMO system comprising $M$ antennas, enabling the base station to simultaneously serve multiple users, potentially from multiple operators. Each operator $q$ is  assigned a slice and a portion $\alpha_q$ of the band, $B$, used by the NH. This band might belong to the NH or might be the concatenation of the bands of the operators. The transmit power budget is allocated equally among all PRBs in a time slot and the power per PRB is $P$.

The $q$th operator serves a set of $U_q$ single antenna users, denoted as $\mathcal{U}_q$ where ${\cal U} = \cup_q \, {\cal U}_q$ is the set of all users and $U  = \sum_q U_q$ is the total number of users in the system.

OFDMA, the dominant multiple access technique in 4G and 5G networks, partitions the available spectrum  into $C$ sub-channels and time into  time-slots. A frame is composed of $T$ time-slots. A pair $(c,t)$ composed of a single sub-channel $c$ and a single  time-slot $t$ makes a Physical Resource Block (PRB). PRBs are the smallest units of resource allocation in the system. In the following, to simplify notation,  PRB $(c,t)$ of a given frame  (where $1 \leq c \leq C$ and $1 < t \leq T$)
is indexed by the integer $n = C(t-1) + c$.

The communication channel linking one of the antennas at the base station to a user is characterized by a combination of small-scale and large-scale fading. The small-scale fading  in PRB $n$ from antenna $m$ to user $u$ is represented by the complex coefficient $h_u^{m,n}$. The large-scale fading channel coefficient, $\rho_u$, represents the impact of power attenuation (due to pathloss) and potential shadowing in the communication link between the base station and the $u$th user. It is taken to be invariant in a frame and the same for all the antennas. Consequently, the complex channel coefficient from the $m$th antenna to the $u$th user on the $n$th PRB is expressed as:
\begin{align}
g_u^{m,n} = \sqrt{\rho_u} h_u^{m,n}, \qquad \forall u \in {\cal U}.
\end{align}

The channel vector between the base-station and user $u$ in PRB $n$ is denoted as $\vec{g}^n_{u} = \big[g^{1,n}_u,g^{2,n}_u, \hdots , g^{M,n}_u\big]^T \in \mathbb{C}^{M\times 1}$, and the channel matrix $\mat{G}^n \in \mathbb{C}^{U\times M}$ for PRB $n$ is given by:
\begin{align}
\mat{G}^n= \big[ \vec{g}^n_{1},\vec{g}^n_{2}, \hdots , \vec{g}^n_{U} \big]^T.
\end{align}

In each PRB $n$, the base-station uses Zero-Forcing Transmission (ZFT) on  a set of selected users $\mathcal{X}_n = \{u_1^n, \ldots, u_{|\mathcal{X}_n|}^n\} \subset {\cal U}$.
For a selected user $u$ in PRB $n$, the ZFT precoding vector is represented as $\vec{\overline{w}}^n_{u}  \in \mathbb{C}^{M}$.
Let $s^n_{u}$ represents a transmitted symbol   sent to user $u$ in PRB $n$ where $\mathbb{E}[|s_{u}^n|^2]=1$, i.e., symbols are normalized to unit average power. 
The vector of transmitted symbols is denoted as $\vec{s}^n = [s^n_{u_1},s^n_{u_2}, \hdots, s^n_{u_{|\mathcal{X}_n|}}]$.
Hence, the power allocated to user $u \in {\cal X}_n$ on PRB $n$ is $p^n_u = ||\vec{\overline{w}}^n_{u}||^2$ and we have
\begin{align}
 \vec{\overline{w}}^n_{u} = \sqrt{p^n_u } \vec{w}^n_{u},
\end{align}
where $ \vec{w}^n_{u} = \vec{\overline{w}}^n_{u} / ||\vec{\overline{w}}^n_{u}||$ is the normalized precoding vector of $\vec{\overline{w}}^n_{u}$.

We define the signal vector transmitted from all base-station antennas on PRB $n$ as: 
\begin{align}
 \overline{\vec{s}}^n = \sum_{u \in \mathcal{X}_n} \vec{\overline{w}}_u^n s^n_u. 
\end{align}
Then the received symbol $y^n_{u}$ at user $u \in \mathcal{X}_n$ is  
\begin{align}
y^n_u 
  =&(\vec{g}^n_{u})^T \overline{\vec{s}}^n + \eta^n_{u} \nonumber \\  
  =& (\vec{g}^n_{u})^T \vec{\overline{w}}^n_{u} s^n_{u} +  
    \sum_{v\in \mathcal{X}_n, v\neq u}(\vec{g}^n_{u})^T \vec{\overline{w}}^n_{v} s^n_{v} + \eta^n_{u},
\end{align}
where $\eta^n_u \sim \mathcal{N}_{\mathcal{C}}(0,\sigma^2)$ is the complex circularly-symmetric additive white Gaussian noise (AWGN).  For ZFT, for each pair of selected users, $u$ and $v$, we have: 
\begin{align}
  (\vec{g}^n_{u})^T \vec{w}^n_{v} = 0,  \qquad u \neq v \land u,v \in \mathcal{X}_n, \forall n
\end{align} and the SNR 
$\gamma^n_u$ for user $u \in \mathcal{X}_n$ at PRB $n$ is then
\begin{align}
 \gamma^n_u = p^n_u||(\vec{g}^n_{u})^T \vec{w}^n_{u}||^2/\sigma^2.
\end{align}

In  PRB $n$, the per PRB power budget $P$ is distributed to all selected users, which can be done by introducing one constraint per PRB, namely,
\begin{align}
\sum_u p_u^n \leq P \quad \forall n.
\end{align}

Finally, we let $f(.)$ be the rate-function that maps an SNR to a rate in bit/s.

%-----------------------------------------------------------------------------------
\section{FGSS Formulation and Offline Solution}\label{sec:DRS}
\subsection{Problem Formulation}
In this section, we present the FGSS formulation that aims to improve spectrum utilization over the static spectrum sharing benchmark, and derive an offline solution. In Section \ref{sec:online_solution}, an online solution to the problem will be presented.
We formulate the FGSS problem for a frame of $N = C T$ PRBs. For ease of notation, the frames are not indexed.

In the benchmark solution, in each time-slot, each network slice $q$ is allocated a fixed share $\alpha_q > 0$ (where $\sum_q \alpha_q = 1$) of the $C$ sub-channels for exclusively serving its users. 
The resource allocation in each network is then decoupled from that of the other networks and each operator aims to allocate resources, via user-selection and power distribution, in a proportional fair (PF) manner, i.e., aims to maximize the geometric mean (geomean) throughput achieved by its users at the end of the frame. 

Formally, the geomean throughput of the users associated with slice $q$ up to PRB $n$ of the  frame ($1 \leq n \leq N$) is 
\begin{equation}
    \Gamma_q(n) = \left(\prod_{u \in \mathcal{U}_q} \sum_{\ell=1}^n r_{u,\ell}^q\right)^{\frac{1}{|\mathcal{U}_q|}}
\end{equation}
where  $r_{u,\ell}^q$ is the rate received by user $u\in \mathcal{U}_q$ in PRB $\ell$. We denote by $\Gamma^b_q(n)$ the geometric mean throughput up to PRB $n$ when the benchmark static sharing is performed. Note that in this case $r_{u,\ell}^q=0$ if PRB $\ell$ has not been allocated to slice $q$.

With FGSS, users from different operators can be multiplexed over any PRB, i.e., user-selection in a PRB is over all users and power distribution is over the set of selected users which may now include users from different operators.
 The resource allocation between networks is now coupled and constraints have to be added to enforce isolation and cross-subsidization avoidance as discussed before. Mathematically, cross-subsidization avoidance is formulated as the constraint that,  at the end of the frame, each operator has been provided the same relative gain with respect to its static spectrum sharing benchmark performance, i.e.,   
\begin{equation}
\frac{\Gamma_q(N)}{\Gamma_q^b(N)} =  \frac{\Gamma_1(N)}{\Gamma_1^b(N)} \quad \forall q \neq 1,  
\end{equation}
where $\Gamma_q(N)$ is the geomean throughput for slice $q$ at the end of the frame  when using FGSS. Similarly, we define isolation as: 
\begin{equation}
    \frac{\Gamma_q(N)}{\Gamma_q^b(N)} \geq 1 \quad \forall q,
\end{equation}
i.e., this constraint ensures that each operator does at least as well  with the proposed FGSS solution as with the benchmark. 

Problem $\textbf{P}_1$ presents the FGSS joint user-selection, ZFT precoding, power distribution, and 
rate selection optimization over a frame of $N$ PRBs, when the benchmark performance metrics $\Gamma^b_q(N)$ are given. Since we are first formulating an offline problem to quantify the performance benefits of FGSS with respect to the benchmark, the $\Gamma^b_q(N)$ can be computed beforehand. How to deal with this in an online setting will be discussed later. 

The objective of Problem $\textbf{P}_1$ is to maximize the relative gains of the slices under the constraints of cross-subsidization avoidance (\ref{eq:CSA_constraint}), and maintaining isolation (\ref{eq:Isolation_constraint}). 
Specifically, Problem $\textbf{P}_1$ maximizes the relative gain of the first slice (with respect to its benchmark performance) since under constraints \eqref{eq:CSA_constraint}, this maximizes the relative gain of all slices. 
The maximization is accomplished by selecting,
for each user $u \in \mathcal{U}_q$, each PRB $n$, and each slice $q$,
the user-selection indicator variables $x_{u,n}^q$,  the ZFT precoder variables $\vec{w}_{u,n}^q$, as  well as the allocated power variables  $p_{u,n}^q$.

Recall that, for  user $u$ of network $q$, the achieved rate $r_{u,n}^q$ in  PRB $n$ can be computed as $f(\gamma_{u,n}^q)$ where $f(.)$ is the rate-function. The SNR $\gamma_{u,n}^q$
is defined in terms of the chosen precoders and the allocated power per user as in (\ref{eq:SNR_computation_constraint}).
Eqs.~(\ref{eq:Power_constraint}) and (\ref{eq:Power_to_selected_only_constraint}) constrain the total distributed power among the selected users from all slices to not exceed the total power $P$ allocated to a given PRB and that power is only assigned to selected users. Finally, constraint (\ref{eq:ZFT_no_interference_constraint}) ensures that the selected precoders cause no inter-user interference, i.e., they are zero-forcing.

Problem $\textbf{P}_1$ is a large mixed-integer non-linear problem that is very hard to solve as is. In the following, we propose a method to derive feasible solutions to this problem that will show the gains that can be achieved by FGSS. 

\begin{optim_prob}

\caption{$\textbf{P}_1$ (FGSS): Joint user-selection, ZFT precoding, power distribution, and rate selection optimization over a frame of $N$ PRBs given benchmark performances $\Gamma^b_q(N)$ and a large positive constant  $D$. }
\begin{subequations}
\begin{align}
&\max_{\substack{\{\Gamma_q(N), x_{u,n}^q, \\ p_{u,n}^q, \vec{w}_{u,n}^q,r_{u,n}^q \} } } \; 
\frac{\Gamma_1(N)}{\Gamma_1^b(N)} \label{eq:slice_gain_obj}
\\
&\text{subject to:} \nonumber \\
&\;\; \frac{\Gamma_q(N)}{\Gamma_q^b(N)} =  \frac{\Gamma_1(N)}{\Gamma_1^b(N)} && \forall q \neq 1 \label{eq:CSA_constraint} 
\\ 
&\;\; \frac{\Gamma_q(N)}{\Gamma_q^b(N)} \geq 1 \label{eq:Isolation_constraint}&& \forall q 
\\ 
&\;\;\Gamma_q(N) = \left(\prod_{u \in \mathcal{U}_q} \sum_{n=1}^N r_{u,n}^q \right)^{\frac{1}{|\mathcal{U}_q|}} \label{eq:GM_definition}&& \forall q 
\\ 
&\;\; r_{u,n}^q = f(\gamma_{u,n}^q)  \label{eq:MCS_selection_constraint}&& \forall u,n,q 
\\ 
&\;\; \gamma_{u,n}^q  = p_{u,n}^q || (\vec{g}_{u,n}^q)^T \vec{w}_{u,n}^q ||^2 /\sigma^2 \label{eq:SNR_computation_constraint} && \forall u,n,q
        \\
&\;\; \sum_q\sum_{u \in \mathcal{U}_q} p_{u,n}^q \leq P 
\label{eq:Power_constraint}&& \forall n 
\\ 
&\;\; p_{u,n}^q \leq x_{u,n}^q P  \label{eq:Power_to_selected_only_constraint} && \forall u,n,q
        \\
&\;\; || (\vec{g}_{v,n}^q)^T \vec{w}_{u,n}^q ||^2 \leq (2 - x_{u,n}^q - x_{v,n}^q ) D \label{eq:ZFT_no_interference_constraint}&&\forall u,v\neq u,n,q 
        \\ 
&\;\; ||\vec{w}_{u,n}^q ||^2 =  1  && \forall u,n,q  
        \\ 
&\;\; x_{u,n}^q \in \{0,1\} \label{eq:last_constraint} && \forall u,n,q
\end{align}
\end{subequations}
\end{optim_prob}
%--------------------------------------------------------------------------------------------------------
\subsection{Solution Technique} 
\label{sec:online_scheme}
%--------------------------------------------------------------------------------------------------------

If the isolation constraint \eqref{eq:Isolation_constraint} were dropped, the relaxed problem would either yield a gain greater than one for all slices, i.e., a feasible solution, or it would be impossible to give a gain greater than one. This is because, according to constraint \eqref{eq:CSA_constraint}, all ratios are equal to that of the first slice, and the latter is the objective to maximize in $\textbf{P}_1$. 
Since the benchmark solution is a feasible solution to $\textbf{P}_1$ that provides a gain of at least one in \eqref{eq:Isolation_constraint}, we relax the constraint as there is no loss in principle to doing so. 
Below, a  heuristic solution is developed to solve $\textbf{P}_1$, and in Section~\ref{Sec:results}, it will be found that the constraint is always satisfied in practice with this heuristic.

To proceed further, we replace the objective in Problem $\textbf{P}_1$ and constraint \eqref{eq:CSA_constraint} with equivalents that will be more suitable for applying the method of Lagrange multipliers. Specifically, we consider the problem
\begin{subequations}
\begin{align}
\max \quad  
 &\frac{1}{Q}\sum_q \log \frac{\Gamma_q(N)}{\Gamma_q^b(N)}
 \label{eq:obj} \\
\text{subject to:} \quad &   
\log \frac{\Gamma_q(N)}{\Gamma_q^b(N)} = \frac{1}{Q}\sum_p \log \frac{\Gamma_p(N)}{\Gamma_p^b(N)} && \forall q  \label{eq:CSA_constraint_log}  \\
&\text{ constraints }\eqref{eq:GM_definition} - \eqref{eq:last_constraint}
\end{align} \label{eq:equiv_prob} 
\end{subequations}
where \eqref{eq:CSA_constraint_log} enforces that all slices  obtain the same gain, and the objective in \eqref{eq:obj} is to maximize this common gain.

Given Lagrange multipliers $\{\nu_q\}$ for the constraints \eqref{eq:CSA_constraint_log}, we can transform problem  \eqref{eq:equiv_prob} into problem
\eqref{prob:before_rewrite} below:
\begin{subequations}
    \begin{align}
     &\max_{\substack{\{\Gamma_q(N), x_{u,n}^q, \\ p_{u,n}^q, \vec{w}_{u,n}^q,r_{u,n}^q \} } }  \; 
     \left\{ \frac{1}{Q}\sum_q \log \frac{\Gamma_q(N)}{\Gamma_q^b(N)}\right. \nonumber \\ 
     &\qquad \qquad +  \left. \nu_q \left(\log \frac{\Gamma_q(N)}{\Gamma_q^b(N)} - \frac{1}{Q}\sum_p \log \frac{\Gamma_p(N)}{\Gamma_p^b(N)}  \right) \right\}
     \\
    &\text{subject to:}  \text{ constraints }\eqref{eq:GM_definition} - \eqref{eq:last_constraint}
    \end{align}    \label{prob:before_rewrite}
\end{subequations}
Specifically, if the  $\{\nu_q\}$ are chosen so that the solution to  \eqref{prob:before_rewrite} results in $\{\Gamma_q(N)\}$ that satisfy \eqref{eq:CSA_constraint_log}, then this solution also solves problem \eqref{eq:equiv_prob}  \cite[Proposition 4.3.4]{bertsekas2016nonlinear}. 

We can then rewrite \eqref{prob:before_rewrite} as: 
\begin{subequations}
\begin{align}
   \max_{\substack{\{\Gamma_q(N), x_{u,n}^q, \\ p_{u,n}^q, \vec{w}_{u,n}^q,r_{u,n}^q \} } } \quad &\; \sum_q \beta_q \log \frac{\Gamma_q(N)}{\Gamma_q^b(N)} \\
   \text{subject to:} \quad &\; \text{ constraints }\eqref{eq:GM_definition} - \eqref{eq:last_constraint},
\end{align} \label{prob:beta_log}
\end{subequations} 
where $\beta_q = \left(\frac{1}{Q} + \nu_q -\frac{\sum_p \nu_p}{Q}\right)$ and $\sum_q \beta_q = 1$.

Thus, to solve the FGSS problem, we should identify the choice of $\{\beta_q\}$ such that the solution of \eqref{prob:beta_log} also satisfies \eqref{eq:CSA_constraint_log}.
To do so, we could solve \eqref{prob:beta_log} for a given a set of $\{\beta_q\}$, then verify 
if constraint \eqref{eq:CSA_constraint_log} is satisfied. If it is, we stop, otherwise we modify the set of $\{\beta_q\}$ to hopefully improve the solution. This iterative method is common practice but unfortunately, \eqref{prob:beta_log} is too complex to solve as is. Instead, per-frame resource management problems on the downlink are typically decomposed into  a sequence of myopic  per PRB problems (thus resulting in a sub-optimal solution) where each such problem takes the past into account. 

Instead of solving a sequence of $N$ per PRB problems for a given set of $\{\beta_q\}$ and then verifying constraint \eqref{eq:CSA_constraint_log} to decide if another iteration on the $\{\beta_q\}$ is necessary, which would be very cumbersome, we propose a solution technique that is less complex and has the significant advantage to be usable in an online setting as will be discussed later. Specifically, 
we start with a guess for the $\{\beta_q\}$ in the first PRB and then update this guess after each PRB based on how far constraint \eqref{eq:CSA_constraint_log} is from being satisfied. It will be shown in the results section that this approach works very well, i.e., the gains are quasi equal at the end of a frame in all cases (i.e., over different realizations and system parameters). We now describe in more details the solution technique, starting with the transformation of \eqref{prob:beta_log} into a sequence of per PRB problems.

First note that the objective to maximize in \eqref{prob:beta_log} can be written as:
\begin{align}
    \max \;
    &\sum_q \beta_q \left( \log \frac{\Gamma_q(1)}{\Gamma_q^b(N)} +
    \sum_{n=2}^N \log \frac{\Gamma_q(n)}{\Gamma_q(n-1)} \right).
\end{align}
This comes from writing $\frac{\Gamma_q(N)}{\Gamma_q^b(N)}=\frac{\Gamma_q(1)}{\Gamma_q^b(N)} \prod_{n=2}^N \frac{\Gamma_q(n)}{\Gamma_q^b(n-1)}$.

We now focus on PRB $1$, ignoring subsequent PRBs and their variables and select an initial set of weights $\{\beta_q\}$, that we call $\{\beta_q^{(1)}\}$. Since $\log (\Gamma_q(1)/ \Gamma_q^b(N)) = \log \Gamma_q(1) - \Gamma_q^b(N)$ and $\Gamma_q^b(N)$ is a constant, the problem to optimize in PRB $1$ then becomes
\begin{align}
\max \sum_q \beta_q^{(1)}  \log \Gamma_q(1).
\end{align}
Based on the solution of this first problem, we compute values for $\{\Gamma_q(1)\}$ and then update the weights (we will discuss how later) to obtain the set $\{\beta_q^{(2)}\}$.

Now, assume that after PRB $n-1$, we have computed $\{\Gamma_q(n-1)\}$ and weights $\{\beta_q^{(n)}\}$. Since $\log (\Gamma_q(n)/ \Gamma_q(n-1)) = \log \Gamma_q(n) - \log \Gamma_q(n-1)$ and $\Gamma_q(n-1)$ is now a constant, the problem we solve in PRB $n$ is then 
\begin{align}   
    \max \;
    &\sum_q \beta_q^{(n)} \log \Gamma_q(n). \label{eq:opt}
\end{align}
Based on the solution of this problem, we compute $\{\Gamma_q(n+1)\}$ and then update the weights to obtain the set $\{\beta_q^{(n+1)}\}$.

The $\beta_q^{(n)}$ are sequentially updated such that constraint \eqref{eq:CSA_constraint_log} holds provided $N$ is large enough. 
For this, we propose: 
\begin{align}
\beta_q^{(n+1)} = \frac{\left[\beta_q^{(n)} - \rho \left( \frac{\Gamma_q(n)}{\Gamma_q^b(N)} - \frac{1}{Q}\sum_p \frac{\Gamma_p(n)}{\Gamma_p^b(N)} \right)\right]^+}{\sum_q \left[\beta_q^{(n)} - \rho \left( \frac{\Gamma_q(n)}{\Gamma_q^b(N)} - \frac{1}{Q}\sum_p \frac{\Gamma_p(n)}{\Gamma_p^b(N)} \right)\right]^+},
\label{eq:update}
 \end{align}
where $\rho > 0$ is a step size parameter and $[x]^+ = \max(0,x)$. The update increases the relative weights $\beta_q^{(n+1)}$ of slices $q$ for which the ratio ${\Gamma_q(n)}/{\Gamma_q^b(N)}$ is below the current target 
$\frac{1}{Q}\sum_p {\Gamma_p(n)}/{\Gamma_p^b(N)}$.
The normalization step ensures that the sum of  $\{\beta_q^{(n+1)}\}$ is kept equal to one.  It should also be noted that although the $\beta_q^{(n)}$'s are updated on a per PRB basis,  they could be updated less frequently. It will be shown in Section~\ref{Sec:results}, that this   update method converges quickly.

Finally, we discuss how to solve the per PRB problem \eqref{eq:opt}. We approximate it as a weighted sum-rate as follows.  Let $R_{u,n}^q$ denote the average per PRB rate up to PRB $n$ for user $u$ of operator $q$, i.e., $R_{u,n}^q = \frac{1}{n} \sum_{\ell=1}^n r_{u,\ell}^q$. The evolution of $R_{u,n}^q$ then follows  
\begin{align}
    R_{u,n}^q = \left(1-\frac{1}{n}\right)R_{u,n-1}^q + \frac{1}{n}r_{u,n}^q. 
    \label{eq:average_rate_recursion_update}
\end{align}
Then, we have
\begin{align}
    &\log \Gamma_q(n) = \log\frac{1}{n}\left(\prod_{u \in \mathcal{U}_q} \left[ (n-1) R_{u,n-1}^q +  r_{u,n}^q\right]\right)^{\frac{1}{|\mathcal{U}_q|}} \nonumber\\
    &= -\log n +\log \Gamma_q(n-1) \nonumber\\
    &\qquad\qquad  +\frac{1}{|\mathcal{U}_q|}\sum_{u\in\mathcal{U}_q}\log\left(1+\frac{r_{u,n}^q}{(n-1) R^q_{u,n-1}}\right)  
    \nonumber\\
    &\approx -\log n + \log \Gamma_q(n-1) + \frac{1}{|\mathcal{U}_q|}\sum_{u\in\mathcal{U}_q}\frac{r_{u,n}^q}{(n-1) R^q_{u,n-1}}
\nonumber 
\end{align}
where, the last step is obtained using $\log(1+x)\approx x$ for small $x$.
Noting that the only non constant term in the last equation is the last sum, we can replace  
\eqref{eq:opt} by the following weighted sum-rate problem for PRB $n$ to be solved sequentially in each PRB, where $\beta_q^{(n)}$ and $R_{u,n}^q$ are   updated according to \eqref{eq:update} and \eqref{eq:average_rate_recursion_update} in each PRB.
\begin{subequations}
    \begin{align}
        \max_{\substack{\{x_{u,n}^q, p_{u,n}^q, \vec{w}_{u,n}^q,r_{u,n}^q \} } } \quad & \sum_q \frac{\beta_q^{(n)}}{|\mathcal{U}_q|} \sum_{u\in\mathcal{U}_q}\frac{r_{u,n}^q}{R^q_{u,n-1}}\\
        \text{subject to:} \quad & \text{ constraints }\eqref{eq:MCS_selection_constraint} - \eqref{eq:last_constraint}.
    \end{align}
    \label{prob:weighted_sum_rate}
\end{subequations}

\vspace{-6pt}
The problem in \eqref{prob:weighted_sum_rate} is a weighted sum-rate MU-mMIMO resource allocation problem that involves all users ${\cal U}$. 
This is a difficult combinatorial problem, and to obtain a good solution, one must solve the rate assignment problem multiple times for different choices of user selections.  To obtain an offline solution against which online solutions can be benchmarked, we adopt the so-called GDAW-s-FD user-selection strategy along with ZF precoding and the water-filling-based power distribution proposed in \cite{hussein2023Operating}.

In summary, the proposed FGSS solution solves a sequence of per PRB  weighted sum-rate problems \eqref{prob:weighted_sum_rate}, updating the average per PRB rates $R_{u,n}^q$  and the weights $\beta_q^{(n)}$  at the end of each PRB. The FGSS solution balances the performance of each slice while maintaining fairness across all slices over $N$ PRBs. The resulting algorithm is given in Algorithm~\ref{algo:lagrangian_solution}.

%\vspace{-6pt}
\subsection{The Benchmark Problem and Solution}
With respect to the  benchmark, there are   $Q$ independent resource allocation sub-problems that can be solved separately. Specifically, each VNO $q$ receives a fixed share $\alpha_q N$ of the $N=CT$ PRBs and independently allocates its resources to its users to maximize
its geometric mean $\Gamma_q^b(N)$. This allocation problem is the exact problem  
studied in \cite{hussein2023Operating}. Thus, we adopt the solution method of \cite{hussein2023Operating} for computing the static benchmark performance.

\begin{algorithm}[t]
\caption{Offline Resource Allocation with Lagrangian Approach}
\label{algo:lagrangian_solution}
\begin{algorithmic}[1]
\State \textbf{Input}: 
\State \quad $\rho$ - step size parameter 
\State \quad $\Gamma_q^b(N)$ - fixed benchmark target
\State \quad $\epsilon > 0$ - a small number
\State \textbf{Initialize}:
\State \quad Set $\beta_q^{(1)} = 1/Q$ for all VNOs $q$
\State \quad Set $R_{u,0}^q = \epsilon$ for all $u,q$
\For {each PRB $n =   1, ..., N$}
    \State \% Performance resource allocation
    \State Assign $r_{u,n}^q$ from solving \eqref{prob:weighted_sum_rate} as in \cite{hussein2023Operating}.
    \label{step:ra}
    \State Update $R_{u,n}^q$ as in \eqref{eq:average_rate_recursion_update}.
    \State Update $\Gamma_q(n) = \prod_{u\in\mathcal{U}_q} \left(R_{u,n}^q\right)^{1/|\mathcal{U}_q|}$
    \State \% Update the Lagrange multipliers   
    \State Compute weights $\beta_q^{(n+1)}$ from \eqref{eq:update}    
    \State Increment $n$ (move to the next PRB)
\EndFor
\end{algorithmic}

\end{algorithm}

We defer the numerical results validating our approach to Section~\ref{Sec:results}. We focus next on providing an online (i.e., real-time) solution for FGSS since the offline results show that it can bring significant gains with respect to the benchmark and hence, providing online solutions is the next challenge.

%===================================================================================
\section{Online Solution for FGSS} \label{sec:online_solution}
%===================================================================================

While Algorithm~\ref{algo:lagrangian_solution} offers a dynamic approach that effectively mitigates cross-subsidization by adjusting weights on a per PRB basis,  it cannot be used as is online for two main reasons. The first is that, Algorithm~\ref{algo:lagrangian_solution} requires the knowledge of the benchmark performance, $\Gamma^b_q(N)$, beforehand (i.e., at the beginning of the frame) which means that we must obtain an accurate \emph{estimate} for it based on some to be determined information.
Second, an efficient online resource allocation method is required to solve  the weighted sum-rate problem in Step~\ref{step:ra} of Algorithm~\ref{algo:lagrangian_solution}.  Although \cite{hussein2023Operating} also proposed a near optimal online resource allocation heuristic, called FDWG, this heuristic was for the equivalent of a single slice using proportional fairness. However, this online heuristic has a user-selection method that was found to be sub-optimal in the case of multiple slices when the objective is to maximize  performance while avoiding cross-subsidization. Hence, we propose below a new (more efficient) user-selection.

In Section~\ref{subsec:gnn_prediction}, we propose a method to  estimate $\Gamma^b_q(N)$ at the start of the frame using a machine learning model based on a Graph Neural Networks (GNN). In Section~\ref{subsec:resource_allocation}, we propose
a user-selection approach that maintains good performance at low complexity when combined with precoding and the power distribution technique developed in \cite{hussein2023Operating}.

%-----------------------------------------------------------------------------------
\subsection{Predicting the Benchmark Performance with GNN} \label{subsec:gnn_prediction}
%-----------------------------------------------------------------------------------

The update equation~\eqref{eq:update} is based on the benchmark performance $\Gamma_q^b(N)$ for all $q$. While in the offline study, we have assumed that the $\{\Gamma_q^b(N)\}$ are known, a method to estimate them is needed. We propose a method that only relies  on the large-scale fading channel coefficients $\rho_u$ of the users (which change rarely compared to the channel vectors), and hence the estimates can be recomputed only once every several frames. 
Specifically, we propose the use of GNNs as they are 
well-suited for online operation since  they are naturally able to handle data with varying numbers of inputs in a way that is invariant to permutations of the input data. 

The objective of the GNN is to accurately estimate the benchmark $\Gamma_q^b(N)$ for a given slice $q$ within a given scenario. 
Thus, we train independent GNN models for each  deployment scenario (e.g., urban macro  with $M$ antennas at the BS, $N$ PRBs in a frame, a cell radius $R$, and total power budget $P$).

We model the GNN as an unweighted fully connected graph, with each node representing a user. 
The node representing user $u$ is assigned i) the large-scale fading coefficient  $\rho_u$, 
ii) the number of PRBs $C_q$ allocated to the slice that user $u$ is associated with, and iii) the total number of users $U_q$ in the slice. It should be noted that since each slice is provided with isolated resources in the benchmark, we generate independent estimates, $\hat{\Gamma}_q^b(N)$, for each slice. All these features are simple to obtain in practice. 

Although the problem is a standard supervised regression problem, and thus the standard loss function is the Mean Squared Error (MSE), we have opted to use the Mean Absolute Percentage Error (MAPE) as the training loss metric. This is motivated by 
the fact that MAPE is inherently more interpretable, providing a direct percentage-wise measure of prediction accuracy. This is critical since the end goal is to avoid cross-subsidization between slices by providing equal gain to each slice.

We train our GNN for each scenario on $K$ realizations, each one corresponding to a single instance of the system parameters used as input features. 
Then, MAPE is defined as the average of the absolute percentage differences between the predicted and true values. Mathematically, it is expressed as:
\begin{align}    
    \text{MAPE} = \frac{1}{K} \sum_{i=1}^{K}  \frac{\left|\Gamma_{i}^b - \hat{\Gamma}_{i}^b  \right|}{\Gamma_{i}^b} \times 100 \% \label{eq:MAPE_definition}
\end{align}
where $\Gamma_{i}^b$ and $\hat{\Gamma}_{i}^b$ represent the true and predicted values of the geometric mean throughput for the benchmark  for the $i$-th realization.

Finally, in order to build an accurate estimator, a grid-search was performed over the GNN model hyper-parameters where a large number of combinations of hyper-parameters  were tried and the best combination was chosen based on its MAPE performance over a separate validation set. 
The considered hyper-parameters were: i) convolution type (e.g., 
Graph Convolutional Network (GCN) \cite{kipf2016semi}, GraphSAGE \cite{hamilton2017inductive}, and Graph Attention Network (GAT) \cite{velivckovic2017graph}), ii) number of hidden layers, iii)  number of hidden channels, iv)  batch size, and v)  learning rate. Specifically, the choice of SAGE convolution type with 7 hidden layers, each with 64 hidden channels, and  the Rectified Linear Unit (ReLU) activation function, was the best or nearly the best for all scenarios. The batch size was set to 1024, and learning rate to $10^{-2}$.

We will see in Section~\ref{Sec:results} that the accuracy of the approach is excellent, i.e., the proposed predictor has a test MAPE score of less than 5\% for a wide-range of scenarios.

\vspace{-10pt}
\subsection{User-selection} \label{subsec:resource_allocation}
%-----------------------------------------------------------------------------------

An efficient practical resource allocation includes user-selection, precoding, power distribution and rate selection.
It aims at striking a balance between low computational complexity and achieving near-optimal performance. 

In \cite{hussein2023Operating}, the proposed online solution included: 1) a simple user-selection approach, and 2) an online algorithm  that computes a near-optimal solution to the joint power-distribution and rate selection problem, given a user set  and assuming ZFT precoding. However, the simple user-selection turned out to be inefficient in a multi-slice scenario.
In the following,  we propose  a novel user-selection algorithm, denoted as ${\sf Select}(K)$, while re-using the online algorithm from \cite{hussein2023Operating} for  joint power-distribution and rate selection. 
${\sf Select}(K)$ initially selects all users from all slices and computes the weighted sum-rate per user after ZFT precoding, power distribution, and  rate selection. Afterwards, the $K$ users with the highest non-zero weighted rates are kept and  precoding, power distribution, and rate selection are recomputed for these users. In Section~\ref{Sec:results}, different values of $K$ will be tried and the value that yields the highest performance will be used when reporting the performance.

%===================================================================================
\section{Numerical Results}\label{Sec:results}
%===================================================================================
\subsection{Rate Function}
\label{sec:ratefunc}
In practical wireless communication systems, the base station is restricted to a set of $L$ modulation and coding schemes (MCS) and a specific MCS  must be chosen for each transmission to a user. To ensure that the system maintains a given target block error rate (BLER), the base station uses a rate function that maps the SNR $\gamma_u$ of user $u$ to the rate $r_u$ that corresponds to the highest feasible MCS that maintains the selected target BLER. The rate function is a non-decreasing piece-wise constant function defined as: 
\begin{align}
    f(\gamma) =  B_c \sum_{l= 1}^{L} e_l \, \mathbbm{1}_{ [\Gamma_l,\Gamma_{l+1})}(\gamma), \label{eq:true_rate}
\end{align}
where $B_c$ denotes the bandwidth of a subchannel, $e_l$ is the spectral efficiency of MCS $l$ measured in bit/s/Hz, and $\mathbbm{1}_{ [\Gamma_l,\Gamma_{l+1})}(\gamma)$ is an indicator function with a value of 1 if $\gamma \in [\Gamma_l,\Gamma_{l+1})$ and zero otherwise. We denote the SNR decoding threshold for MCS $l$ as $\Gamma_l$ where $\Gamma_{L+1} = \infty$. An example rate function is presented in Table~\ref{tab:mcs-table_v2} \cite{Yue2020}. It should be noted that if the SNR is less than $\Gamma_1$ then the received rate is zero. i.e., $f(\gamma) = 0,  ~\forall \gamma< \Gamma_1$.

%-----------------------------------------------------------------------------------
\subsection{Simulation Settings}
%-----------------------------------------------------------------------------------

In order to evaluate and compare the proposed solution, extensive numerical computations are carried out
for different 3GPP scenarios, e.g., urban macro (UMa), rural macro (RMa) and urban micro (UMi).  To each of these scenarios, corresponds a specific channel model. They differ in terms of large-scale fading, small-scale fading, LOS probability, etc.

First note that a Line-of-Sight (LOS) component is not always present in practice, and hence the model has a location-dependent LOS probability. Following 3GPP \cite{3gpp.38.901}, given a distance $d_{2D}^{u}$ in the X-Y plane between user $u$ and the base station, the probability that user $u$ observes a LOS condition is given by:
\begin{align*}
\text{Pr}_{LOS}^{u} = \begin{cases}
    1, & d_{2D}^u \leq d_{min}, \\
    \frac{\Theta_1}{d_{2D}^u} + \big(1-\frac{\Theta_1}{d_{2D}^u}\big) \exp\big( - \frac{d_{2D}^u}{\Theta_2}\big) &  d_{2D}^u > d_{min},
\end{cases}
\end{align*}
where the constants, $d_{min},\Theta_1$, and $\Theta_2$ vary according to the scenario as defined by \cite[Table 7.4.2-1]{3gpp.38.901}. 

We now describe the large-scale fading 
part of the channel model, common to the three scenarios (with different values of the parameters). 
The large-scale fading coefficient, $\rho_u$, for user $u$ is modeled as a log-normal random variable, with a mean path-loss of
\begin{align}
    \mu_u  = \Theta_3 + \Theta_4 \log(d_{3D}^u) + \Theta_5 \log(f_c) \quad \text{(dB)},
\end{align}
and a standard deviation $\Theta_6$, representing 
random shadowing. The model coefficients $(\Theta_3, \Theta_4, \Theta_5, \Theta_6)$ are given in \cite[Table 7.4.1-1]{3gpp.38.901} for the different scenarios and depend on the user's LOS condition.

For small-scale fading, we adopt a Rician fading model. We assume that $\vec{g}_u^n$, the channel vector for user $u$ in PRB $n$, is composed of a Line-of-Sight (LOS) component and a component resulting from multi-path propagation,  i.e.,

\begin{align}
    \vec{g}_u^n = \sqrt{\frac{\rho_u}{1+\kappa_u}} \vec{h}^\text{G}_{u,n} \mat{R}^{1/2} +  \sqrt{\frac{\kappa_u \rho_u}{1+\kappa_u}}  \vec{h}^{\text{LOS}}_{u}, \qquad \forall u,
\end{align}
where $\kappa_u$ is the Rician factor and $\vec{h}^{\text{LOS}}_{u}$ represents the LOS component for the $u$th user. $\vec{h}^\text{G}_{u,n} $ is a vector whose entries are zero-mean i.i.d.~complex Gaussian random variables, and $\mat{R}$ is the transmit antenna correlation matrix assumed to be the same for every user. 
For the Ricean factors $\kappa_u$, we draw them from a Gaussian distribution that depends on the deployment scenario following the 3GPP model \cite[Table 7.5-6]{3gpp.38.901}. 
For users without a LOS component, we set $\kappa_u =0$ and their channels reduce to a correlated Rayleigh fading model.

If a LOS component exists for user $u$, the LOS component $h^{\text{LOS},m}_{u}$ for the $m$-th antenna, is given by
\begin{align}
    h^{\text{LOS},m}_{u} = \exp{(2\pi j  || \vec{p}_u - \vec{a}_m ||/\lambda)} 
\end{align}
where $ \vec{p}_u \in \mathbb{R}^3$ is a vector denoting the position of user $u$,  $\vec{a}_m$ denotes the position of the $m$-th BS antenna, and $\lambda$ is the wavelength.

\begin{table*}[t]
\caption{Available rates and the corresponding SNR thresholds \cite{Yue2020}.
}
\label{tab:mcs-table_v2}
\resizebox{\textwidth}{!}{%
\begin{tabular}{|l|l|l|l|l|l|l|l|l|l|l|l|l|l|l|l|}
\hline
SNR Threshold (dB) $\Gamma_l$     & -6.82 & -3.44 & -0.53 & 3.79 & 5.80 & 8.08 & 9.76 & 11.72 & 13.49 & 15.87 & 17.73 & 19.50 & 21.30 & 23.51 & 25.15 \\ \hline
Efficiency (bits/s/Hz) $e_l$ & 0.15 & 0.38 & 0.88 & 1.48 & 1.91 & 2.41 & 2.73 & 3.32 & 3.90 & 4.52 & 5.12 & 5.55 & 6.23 & 6.91 & 7.41  \\ \hline
\end{tabular}%
}\end{table*}

The BS is assumed to be equipped with a uniform planar array (UPA) featuring $\sqrt{M}$ antennas in both vertical and horizontal directions, uniformly spaced at half a wavelength. Consistent with the approach in  \cite{ying2014kronecker,loyka2001channel}, we adopt an exponential correlation model for the antennas. In this model, the $(i,j)$-th entry of the correlation matrix $\mat{R}$  is defined as 
\begin{align}
    [\mat{R}]_{i,j} = \psi^{|v(i)-v(j)|+|h(i)-h(j)|}, \quad \forall i,j = 1,..., M,
\end{align}
where $0\leq \psi \leq 1$ denotes the correlation coefficient between adjacent antennas; $v(m)$ and $h(m)$ are the vertical and horizontal positions of the $m$-th antenna within the UPA.
  
The values selected for the remaining system parameters are presented in Table~\ref{tab:Sim_parameters}.
Users locations are drawn uniformly at random in the cell and the fading coefficients are computed as described earlier in this section. 
The power levels are chosen such that 99\% of users, if scheduled alone, have a single user MIMO SNR of at least -3.44 dB which allows them to successfully decode the second lowest MCS defined in Table~\ref{tab:mcs-table_v2}. If we sample a user that cannot decode the lowest MCS, i.e., has a single user MIMO SNR less than -6.82 dB, then we redraw the position of the user.

\begin{table}[t]
\caption{System simulation parameters.}
\centering
\begin{tabular}{|c|c|c|}
\hline
\textbf{Parameter} & \textbf{Symbol} & \textbf{Value} \\ \hline
Carrier frequency       & $f_c$      & 2.5 GHz    \\ \hline
    &    & 5 W (UMA) \\
                Total available power           &    $P$               & 5 W (RMA) \\
                          &                   & 2.5 W (UMi)\\ \hline
Number of antennas      & $M$       & 100      \\ \hline
Number of channels      & $C$       & 100      \\ \hline
PRBs in a frame & $N$ & 2000 \\ \hline  
                     &        & 400 m (UMa)    \\ 
Cell radius          &     $R$      & 1000 m (RMa)    \\ 
                     &           & 200 m (UMi)    \\ \hline                     
PRB bandwidth       & $B_c$      & 180 KHz      \\ \hline
Number of realizations    & $N_{r}$   &100   \\ \hline
Correlation coefficient & $\psi$ & 0.01 \\ \hline
Gradient step size & $\rho$ & 0.01 \\ \hline
 &  & $\mathcal{N}(9,3.5^2)$ (UMa) \\ 
  Rician factor            &         $\kappa_u$     & $\mathcal{N}(7,4^2)$  (RMa) \\
              &              & $\mathcal{N}(9,5^2)$  (UMi) \\              \hline
\end{tabular}
\label{tab:Sim_parameters}
\vskip -8pt
\end{table}

%------------------------------------------------------------------------------------
\subsection{The Offline Study}
%------------------------------------------------------------------------------------
In this section, we compute the performance of the proposed solution  presented in Algorithm~\ref{algo:lagrangian_solution} and compare its performance to the fixed-share benchmark solution. 
Here, 
the benchmark performances $\Gamma_q^b(N)$ are first computed numerically, at the beginning of a frame, and taken as known when needed in \eqref{eq:update}. In Sections~\ref{subsec:GNN_result} and \ref{subsec:online}, GNNs will be used to estimate the performance of the benchmark.

\begin{figure}[t]
    \centering
    \includegraphics[width = \columnwidth]{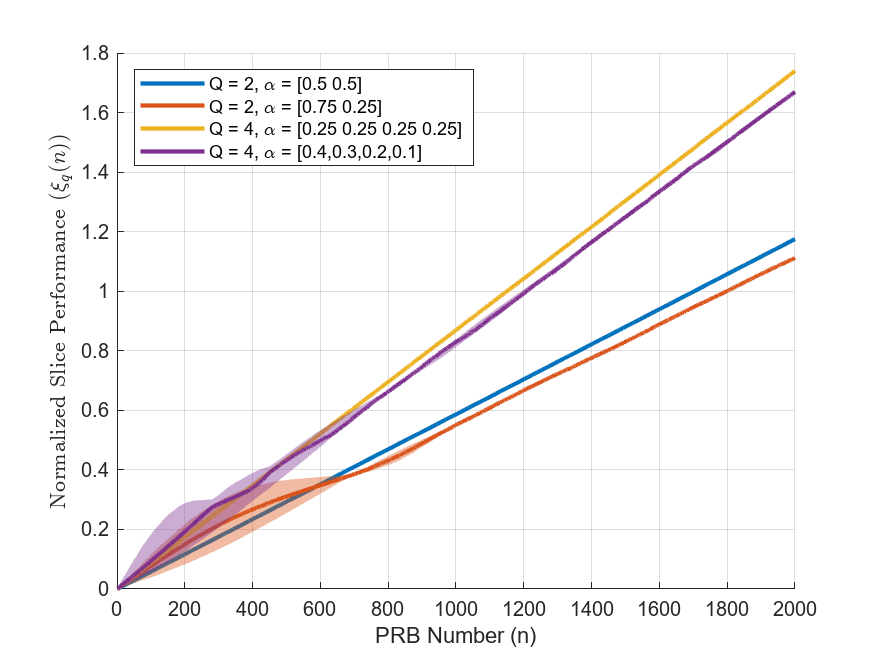}
    \caption{
    Sample paths of the normalized performance achieved with Algorithm~\ref{algo:lagrangian_solution} for configurations (a), (b), (c) and (d)  with $M=100$ antennas over $N = 2000$ PRBs for the UMa scenario. 
    }  
    \label{fig:All_convergence_gain_cases}
\end{figure}

We first analyze the convergence behaviour of the  proposed algorithm by observing the gain as a function of the PRB number $n$ for a single realization. Specifically, we plot the normalized performance per slice $q$, defined as $ \xi_q(n) = \Gamma_q(n)/\Gamma_q^b(N)$ versus the PRB index $n$ for $1 
\leq n \leq N$ where $N$ is the number of PRBs in a frame.  
In Fig.~\ref{fig:All_convergence_gain_cases}, we consider four different configurations for an UMa scenario with $M=100$ antennas:
\begin{enumerate}
\item[(a)] $Q = 2$ identical slices, i.e., $\alpha_q = 0.5$ and $U_q = 15$.
\item[(b)] $Q = 2$ slices with $\alpha_1 = 0.25$ and $\alpha_2 = 0.75$, and again $U_q = 15$ for each slice.
\item[(c)] $Q = 4$ identical slices, with $\alpha_q = 0.25$ and $U_q  = 10$. 
\item[(d)] $Q = 4$ slices with $\alpha_1 = 0.4$, $\alpha_2 = 0.3$, $\alpha_3 = 0.2$ and $\alpha_4 = 0.1$, and $U_q = 10$ for each slice.
\end{enumerate}  
A line is used to show the median of the slice gains
with a surrounding patch 
representing the maximum and minimum gain per slice at a given PRB. 
From Fig.~\ref{fig:All_convergence_gain_cases}, we see that that the patch narrows and the gains of all slices become identical and linear with respect to $n$ almost immediately for the configurations with identical slices (i.e., configurations (a) and (c)). For configurations with different fixed shares $\alpha_q$ (i.e., configurations (b) and (d)), we see that the algorithm results in the slices having equal gain after $n=700$ and $n=800$ PRBs, respectively. Moreover, the proposed algorithm consumes fewer PRBs to achieve similar performance to the fixed-share benchmark, i.e., normalized gains of $\xi_q(n)=1$ are achieved for $n < N$, resulting in significant bandwidth savings of  14.9\%, 10.1\%, 42.5\%, and 39.03\% for configurations (a),(b), (c), and (d), respectively.

Fig.~\ref{fig:CDF_gain_Q_4_equal} shows, for the case of $Q = 4$ identical slices, i.e., $\alpha_q = 0.25 ~\forall q$, the cumulative distribution function of the normalized performance, $\xi_q(N)$, at the end of the frame over 100 realizations. From the figure, we see that $\xi_q(N) > 1$ in all cases, indicating that there is always a positive gain. Also, the median gain of the proposed algorithm over the benchmark ranges from 24.4\% to 81.3\% (depending on the scenario and the number of users), 
and the gain  is at least 148\% in 25\% 
of the realizations for the UMi case with $U_q = 10$.

For the case of $Q=4$ non-identical slices with $(\alpha_1, \alpha_2, \alpha_3, \alpha_4) = (0.4,0.3,0.2,0.1)$ (not shown for brevity), the median slice gain ranges from 17.7\% to 67.9\% (depending on the scenario and the number of users), {and the gain is at least 114\% for 25\% 
of the realizations for the UMi scenario with $U_q=10$.
Additionally, for the case of two non-identical slices with shares $\alpha_1 = 0.75$ and $\alpha_2 = 0.25$ (not shown for brevity), the median slice gain 
varies from 5\% to 18.6\% (depending on the scenario and the number of users), and is at least 28\% for 25\%  
of the realizations for the UMi scenario with $U_q=15$.
From these results, the gains are larger when the number of slices is larger. The gains are also larger when the number of users per slice is not too large (as this allows for more multiplexing opportunities).

\begin{figure}[t]
    \centering
    \includegraphics[width = \columnwidth]{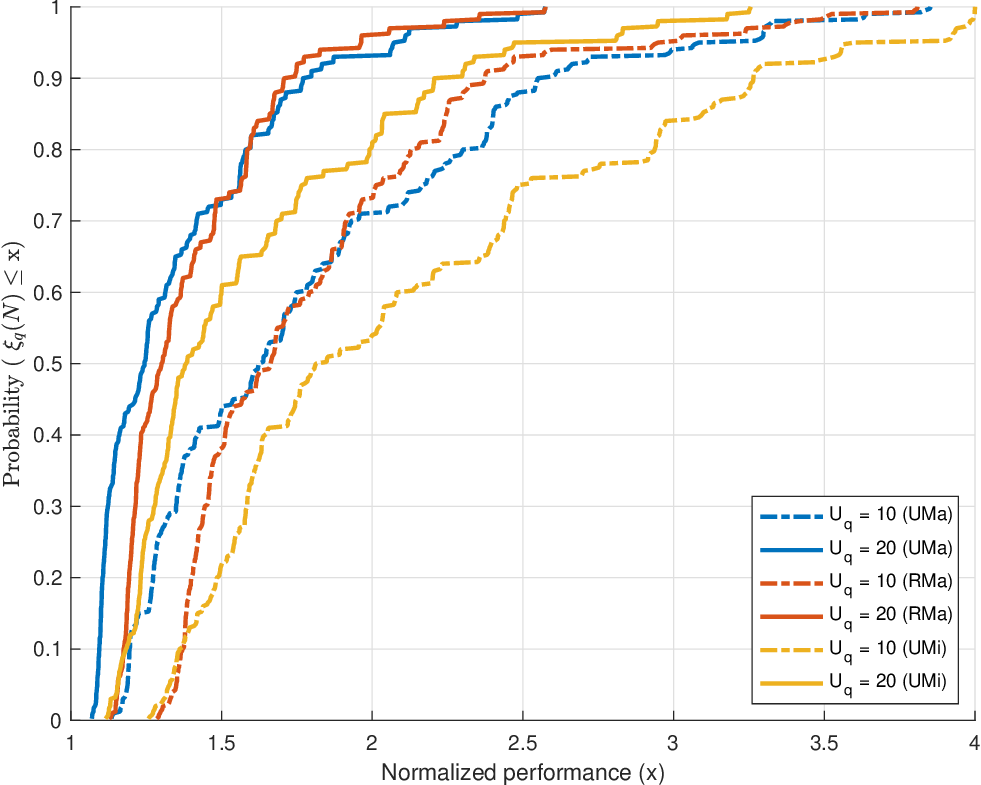}
    \caption{Cumulative distribution function of the normalized performance per slice at the end of the frame $\xi_q(N)$ for $Q=4$ slices, assuming $\alpha_q = 0.25 ~\forall q$, for different values of numbers of users per slice $U_q$ and different deployment scenarios.}
    \label{fig:CDF_gain_Q_4_equal}
\end{figure}

%------------------------------------------------------------------------------------
\subsection{GNN Prediction Accuracy}
\label{subsec:GNN_result}
%------------------------------------------------------------------------------------

We now analyze the benchmark prediction component of the online solution by investigating the quality of the fixed sharing performance estimates generated by the proposed GNN estimator. The benchmark performance is computed for each slice independently. Hence, in the following we focus on a slice and omit the index $q$.

For training purposes, the benchmark performance of 
about 20,000 
realizations with different slice shares in the range $\alpha \in [0.1, 0.75]$ and with 10 to 75 users per slice have been collected.
The GNN model was trained in various deployment scenarios, such as urban macro, rural macro and urban micro. For testing, 2520 additional realizations were generated. Recall that the chosen metric for assessing the model performance is the Mean Absolute Percentage Error (MAPE) of the predicted geometric mean defined in \eqref{eq:MAPE_definition}. 

In Fig.~\ref{fig:Mape_vs_alpha_U_20__M_100},  the MAPE testing score for various values of a slice's share $\alpha$ is plotted for the case of 20 users  and 100 antennas at the base-station. Clearly,  the proposed GNN predictor that only takes in the large-scale fading estimates as well as the static system parameters is capable of predicting the benchmark geometric mean throughput with a MAPE score of at most 5\%. 
In Fig.~\ref{fig:Mape_vs_U_alpha__p2__M_100}, the MAPE is shown for different numbers of users $U$ for $M=100$ antennas and a slice share of $\alpha = 0.2$.  Here, the MAPE error is always less than 6\%.

\begin{figure}[t]
    \centering
    \includegraphics[width = \columnwidth]{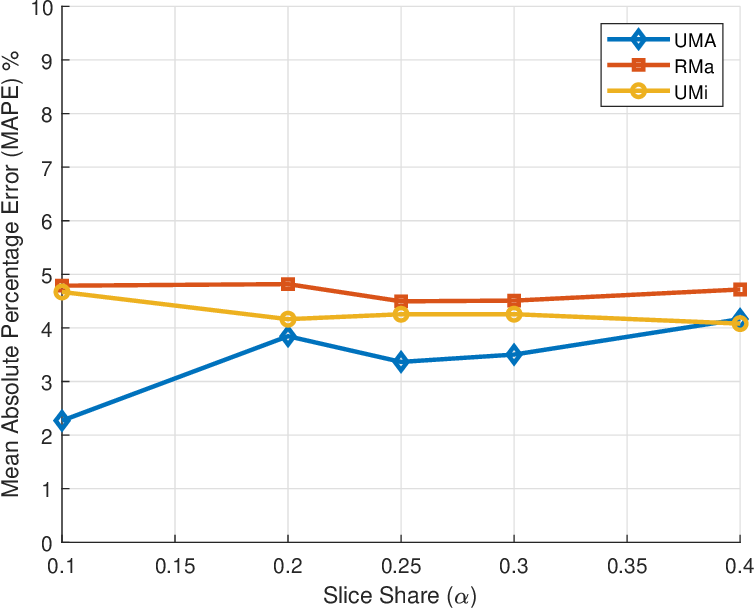}
    \caption{MAPE testing score vs. slice resource share ($\alpha$) for different deployment scenarios when $U=20$ and $M=100$. 
    }
    \label{fig:Mape_vs_alpha_U_20__M_100}
\end{figure}

\begin{figure}[t]
    \centering
    \includegraphics[width = \columnwidth]{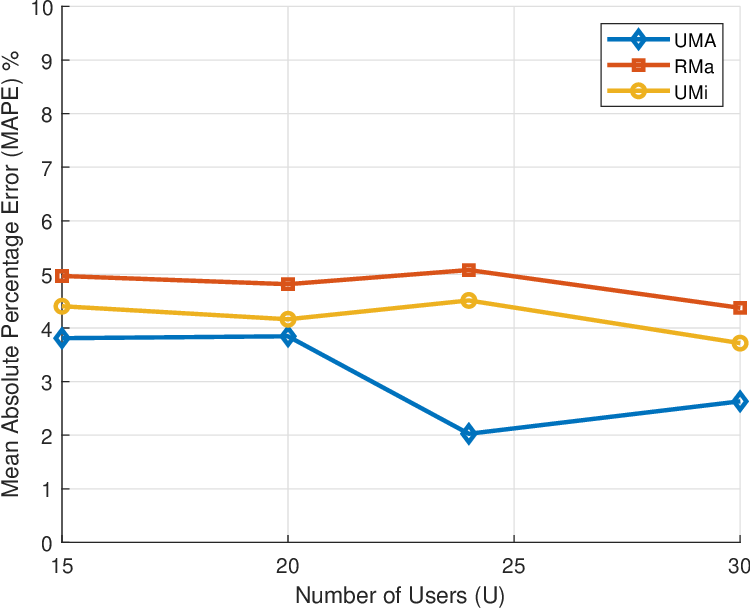}
    \caption{MAPE testing score vs. number of users ($U$) for different deployment scenarios  when $\alpha=0.20$ and $M=100$. 
    }
    \label{fig:Mape_vs_U_alpha__p2__M_100}
\end{figure}

With these good results, we now proceed to analyze the performance of the complete online algorithm.

%------------------------------------------------------------------------------------
\subsection{Online Performance}
\label{subsec:online}
%------------------------------------------------------------------------------------
Recall that the online algorithm is based on two building blocks. The first is the benchmark performance estimator that was studied in the previous subsection, and the second is 
the online user-selection heuristic, ${\sf Select}(K)$, proposed in Section \ref{subsec:resource_allocation}. 
To validate the performance of ${\sf Select}(K)$, 
 we use the exact benchmark performance for computing the normalized gain  of the online algorithm with different choices of $K$. 
Specifically, we consider an urban macro-cell deployment with $Q=3$ identical slices, each with an equal share $\alpha_q = 1/3$ and $30$ users per slice. Fig.~\ref{fig:K_sweep_UMa_S_3_Us_30_30_30_alpha_s_0.33_0.33_0.33_M_100} shows the normalized gain  in geometric mean throughput, $\xi_q(N)$, achieved by i) the offline GDAW-s-FD user-selection method of \cite{hussein2023Operating}, 
and ii) the proposed online user-selection ${\sf Select}(K)$ versus $K$. 

From Fig.~\ref{fig:K_sweep_UMa_S_3_Us_30_30_30_alpha_s_0.33_0.33_0.33_M_100}, for a suitable choice of $K$, the performance of the proposed online scheme is nearly as good as the offline scheme. Specifically, by choosing $K = 30$ for this particular use case, i.e., keeping a third of the available users, at most 6\% of the  offline gain is lost.
However, if $K$ is not chosen properly, the performance drop is as high as 24\%. A similar trend is found for the case where there are $Q = 4$ identical slices, each serving 20 users (the corresponding figure is omitted for brevity).  There, for the best selection of $K$ the gap to the offline scheme is again 6\% and if $K$ is not selected carefully, the performance can degrade by up to 15\% or more. These results are in line with previous findings that emphasize the importance of limiting the number of selected users \cite{hussein2023Operating,bjornson2015massive}.

\begin{figure}[t]
    \centering
    \includegraphics[width = \columnwidth]{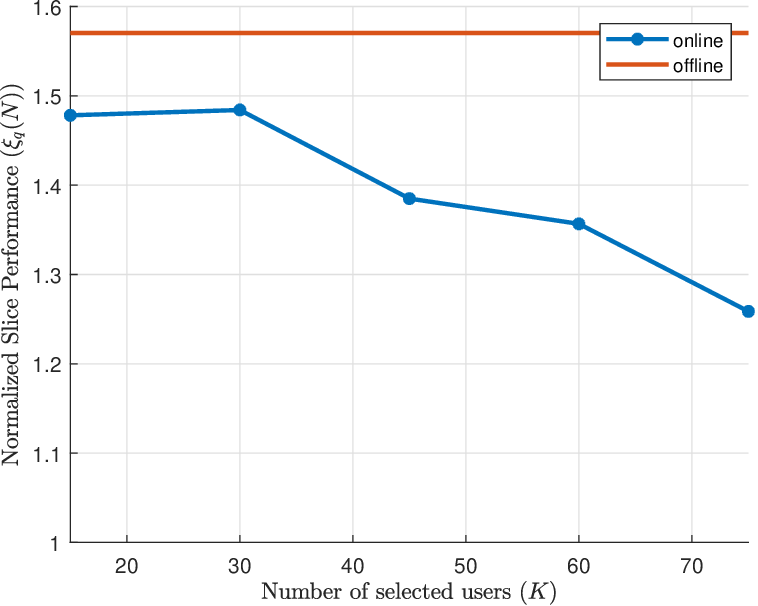}
    \caption{Average normalized performance $\xi_q(N)$ of online ${\sf Select}(K)$ and offline GDAW-s-FD scheme of \cite{hussein2023Operating}
    vs. the number of selected users $K$ for the case of $Q=3$ identical slices with $\alpha_q = 1/3$, $U_q = 30$ and $M=100$ antennas for an urban macro cell deployment. 
    }
    \label{fig:K_sweep_UMa_S_3_Us_30_30_30_alpha_s_0.33_0.33_0.33_M_100}
\end{figure}

Next, we investigate the performance of the  proposed online scheme with both the online user-selection ${\sf Select}(K)$ as well as the GNN estimates
$\hat{\Gamma}_{q}^f(N)$ of the benchmark geometric mean throughput. We have verified (not shown for brevity) that the impact of not having a perfect estimate on the online algorithm is minimal, i.e., the difference in the median gain than 2.8\%. It should be noted that in the results below, the parameter $K$ in the ${\sf Select} (K)$ user-selection algorithm  was chosen via an extensive simulation campaign per deployment scenario. For each scenario,  different configurations (i.e., number of slices, slice shares, and number of users per slice) were tried and for each configuration, many realizations were generated. For each realization, the performance was evaluated for different values of $K$ in the range $1$ to 
$U$, where $U$ is the number of users.

In Fig.~\ref{fig:Offline_vs_Online_ML_S_4_v2.eps}, the statistics of the  slice gain, $\frac{1}{Q}\sum_q (\xi_q(N)-1)\times 100\%$  for the online solution with respect to an online version of the benchmark are shown (they are computed  on 100 realizations for each combination of deployment scenario, slice weight configuration, and number of users per slices $(U_q = 10, 15, 20)$). Each boxplot depicts the median as the central line, the interquartile range (IQR) as the box edges, while the whiskers span the 10th to the 90th percentiles of the 100 realizations.

The first remark is that high gains can be obtained with the online solution. For example, gains higher than 160\% can be obtained for 25\% of the realizations in the case of UMi and UMa with $10$ users per identical slice. Also, as the number of users per slice is increased from $U_q = 10$ to $U_q=20$, the median slice gain decreases across all deployment scenarios. For instance, in the UMi scenario with equal slice shares, the median gain drops from 58\% at $U_q=10$ to 37\% at $U_q = 20$.

\begin{figure}[t]
    \centering    \includegraphics[width=\columnwidth]{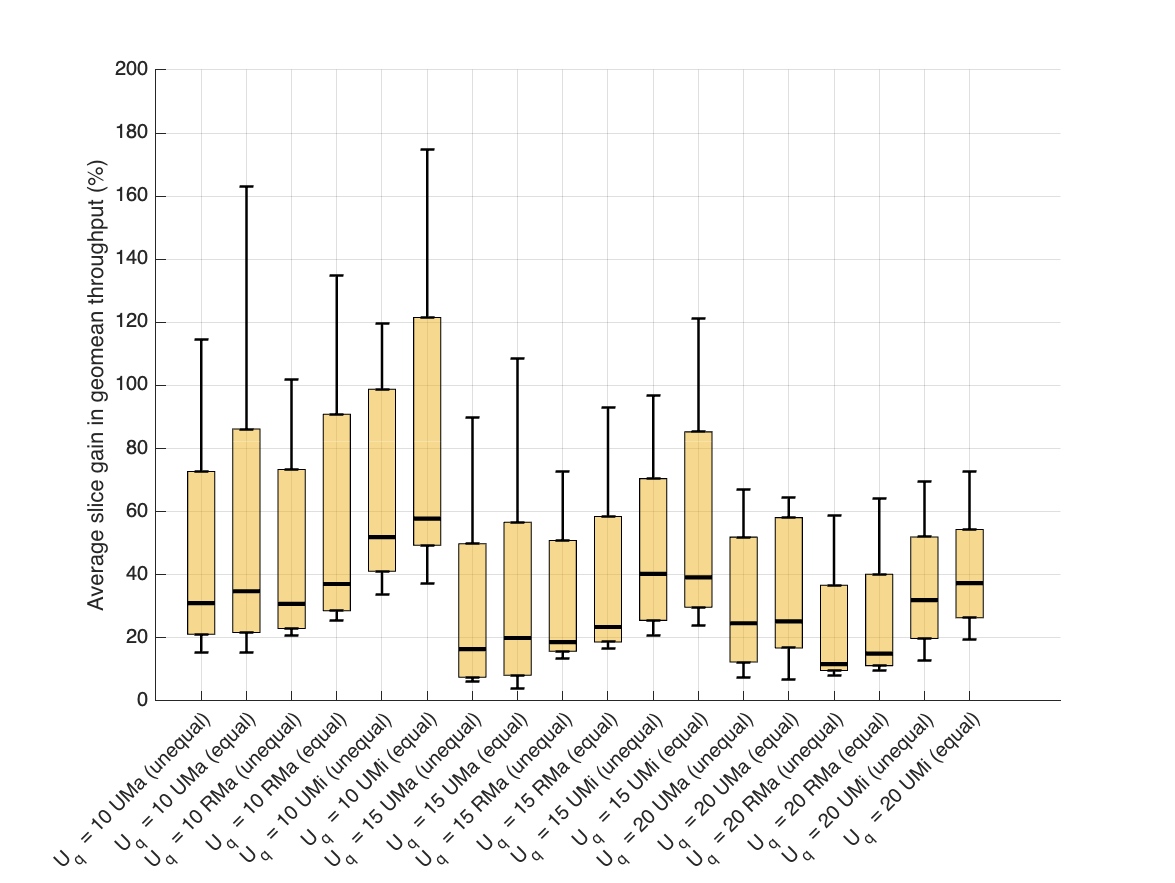}
    \caption{Slice gain statistics for the proposed online solution for Q = 4 slices across different scenarios and numbers of users.}
    \label{fig:Offline_vs_Online_ML_S_4_v2.eps}
\end{figure}

\vspace{-10pt}
\subsection{Complexity of the Online Solution}
The proposed algorithm has a per PRB complexity of $\mathcal{O}(M^2 U + U^2 L )$. The first term corresponds to the first precoding computation for all $U$ users, which is of complexity $\mathcal{O}(M^2 U)$. The second term $\mathcal{O}(U^2L)$ corresponds to the complexity of the online algorithm  for joint power-distribution and MCS selection  \cite{hussein2023Operating}. 
Since only two user-sets are considered per PRB, the worst-case complexity scales only as a function of the complexity of computing precoding and power-distribution.

%===================================================================================
\section{Conclusion }\label{sec:conclusion}
%===================================================================================

In this work, we have analyzed the problem of Fine Grained Spectrum Sharing (FGSS) in massive MIMO neutral host environments. Our framework provides isolation and addresses the risk of cross-subsidization by ensuring equal performance gains for each VNO, irrespective of their resource share. Towards this end, we formulated the FGSS problem as a Mixed Integer Non-Linear Program (MINLP) that aims to maximize the overall sharing gain while guaranteeing isolation and cross-subsidization avoidance. Our proposed Lagrangian approach dynamically adjusts the scheduler weights, achieving equal geometric mean throughput for each VNO relative to its benchmark performance.

To enable practical, real-time, implementation, we introduced two innovations. First, we designed a Graph Neural Network (GNN) estimator that predicts the geometric mean throughput of a slice that is allocated a static share of the spectrum. This throughput is needed  to ensure isolation and avoid cross-subsidization. The GNN estimator 
is trained for a given deployment scenario and has as inputs key parameters such as slice share and the large-scale fading coefficients of the users in the slice.
This method achieves  accurate predictions with a maximum of 5\% mean absolute percentage error and offers a robust and scalable solution for online FGSS adaptation. 
Second, we introduced an online user-selection algorithm, ${\sf Select}(K)$, that has similar performance to that of the offline algorithm introduced in \cite{hussein2023Operating}. ${\sf Select}(K)$ is low complexity and it significantly improves, in the multi-slice scenario, the performance of FDWG, the online algorithm introduced in  \cite{hussein2023Operating}. Based on our extensive evaluations across various slice configurations, deployment scenarios and system parameters,  the proposed real-time scheme that combines ${\sf Select}(K)$ with the GNN predictor is at most 9\% from the offline approach with perfect estimation of the benchmark performance.

The key take-away from our investigation is that the  FGSS framework provides significant gains when compared to the static sharing benchmark, especially for cases with more than two slices.
By enabling dynamic spectrum sharing that maintains isolation and avoids cross-subsidization, FGSS ensures fair and optimized resource distribution across slices, even when slices have different resource shares. For online operation, FGSS relies on the developed GNN to estimate the benchmark performance based on limited information. This GNN is surprisingly accurate. 
Altogether, this work opens new possibilities for collaborative and optimized wireless systems, setting a strong foundation for future advancements in network resource sharing and efficiency.

\bibliographystyle{IEEEtran}
\bibliography{IEEEabriv,Refs}

\end{document}